\documentclass[pra,amsmath,amsfonts,superscriptaddress,showpacs,twocolumn]{revtex4-1}
\usepackage{graphicx}
\usepackage{dcolumn}
\usepackage{subfigure}
\usepackage{bm}
\usepackage{amsmath,amssymb}
\usepackage{color,soul}
\usepackage{xcolor}	
\usepackage{bbold}

\newtheorem{theorem}{Theorem}

\newtheorem{proposition}[theorem]{Proposition}

\newenvironment{proof}[1][Proof]{\begin{trivlist}
\item[\hskip \labelsep {\bfseries #1}]}{\end{trivlist}}

\newcommand{\ket}[1]{|#1\rangle}
\newcommand{\bra}[1]{\langle #1|}

\newcommand{\Id}{{\mathbb 1}}
\newcommand{\e}{{\mathrm e}}

\newcommand{\T}{{\mathrm t}}

\newcommand{\dif}{{\mathrm d}}
\newcommand{\Tr}{{\mathrm {Tr}}}

\begin{document}

\title{Quantum Fisher information matrix  for unitary processes:  closed   relation  for $SU(2)$}

\author{Mohammad Javad Shemshadi}
 \email{mohamadjavad.shemshadi@mail.um.ac.ir}
\affiliation{Department of Physics, Ferdowsi University of Mashhad, Mashhad, Iran}
\author{Seyed Javad Akhtarshenas}
 \email{akhtarshenas@um.ac.ir}
\affiliation{Department of Physics, Ferdowsi University of Mashhad, Mashhad, Iran}

\begin{abstract}
Quantum Fisher information plays a central role in the field of quantum metrology.  In this paper we study the problem of quantum Fisher information of unitary processes. Associated to each parameter $\theta_i$ of unitary process $U(\boldsymbol{\theta})$, there exists a unique Hermitian matrix $M_{\theta_i}=i(U^\dagger\partial_{\theta_i} U)$. Except for some simple cases, such as  when the parameter under estimation is an overall multiplicative factor  in the Hamiltonian, calculation of these matrices is not an easy task to treat even for estimating a single parameter of qubit systems. Using the Bloch vector $\boldsymbol{m}_{\theta_i}$, corresponding to each matrix  $M_{\theta_i}$,    we find a closed relation for the quantum Fisher information matrix of the  $SU(2)$ processes for an arbitrary number of  estimation parameters and an arbitrary initial state.  We extend our results and present  an explicit  relation for each vector $\boldsymbol{m}_{\theta_i}$  for  a general Hamiltonian with arbitrary parametrization. We illustrate our results by obtaining the quantum Fisher information matrix of the so-called angle-axis parameters of a general $SU(2)$ process.
Using  a linear  transformation between two different parameter spaces of a unitary process, we provide a way to move from quantum Fisher information of a unitary process  in a given parametrization to the one of the other parametrization. Knowing  this linear transformation    enables one to calculate the quantum Fisher information of a composite unitary process, i.e. a unitary  process  resulted from successive action of some simple unitary processes.     We apply this  method for a   spin-half system and obtain the quantum Fisher matrix of the coset parameters in terms of the  one of the angle-axis parameters.
\end{abstract}


\maketitle

\section{Introduction}
\label{s1}
Estimation theory is an important topic in different areas of  physics.  Quantum metrology tries to improve estimation precision by using quantum strategy such as entanglement
 \cite{GiovannettiScience2004,GiovannettiNatPhot2011,HuangAnnual2014} and discord \cite{GeorgescuNat2014,GiordaPRL2010,OllivierPRL2001}. Many applications of quantum metrology have been found, such as gravitational radiation \cite{BraginskyPLA2004,AdhikariRMP2014,McGuirkPRA2002}, quantum frequency standards \cite{SantarelliPRL1999,BollingerPRA1996,HuelgaPRL1997}, quantum imaging \cite{TsangPRL2009,GiovannettiPRA2009,BridaNatPhot2010}, and atomic clocks \cite{BuzekPRL1999,AndrePRL2004,LouchetNJP2010,BorregaardPRL2013,KesslerPRL2014}. Estimation precision in quantum metrology is described by the Cramer-Rao inequality \cite{HolevoBook1982,Helstrom1976,BraunsteinPRL1994,BraunsteinAP1996,PetzJPA2002,PetzBook2008}
 \begin{eqnarray}
 \delta\theta=\sqrt{\textrm{Var}(\hat{\theta})}\geq\frac{1}{\sqrt{NF_\theta}},
 \end{eqnarray}
where lower bound is related to the inverse of the quantum Fisher information. The estimation precision for separable states is bounded by the standard quantum limit  $\Delta\theta\sim\frac{1}{\sqrt{N}}$, whereas  for  the maximally entangled states, GHZ and NOON states, it  is bounded by the Heisenberg limit   $\Delta\theta\sim\frac{1}{N}$  \citep{GiovannettiPRL2006,LeePRL2006,PezzePRL2009}. In general, there are three stages in quantum metrology: the first is the preparation of  the input state, the  so-called probe state.  In the second stage   the input state is encoded  with an unknown  parameter $\theta$.  Finally, the third stage is information extraction, carried  out by measuring on the output states. Fisher information is at the heart of metrology and gives us knowledge about the unknown parameters from the probability distribution. It  can be obtained directly from its definition $F_{\theta}=\sum_x P_{\theta}(x)[\partial_{\theta}\ln P_{\theta}(x)]^{2}$, for discrete outcomes $x$  \cite{Fisher1925}, where $P_{\theta}(x)$ is the probability distribution  obtained by measuring the encoded probe states. The maximum of $F_{\theta}$ over all possible measurements is the so-called quantum Fisher information (QFI). Quantum Fisher information is related to the Bures  \cite{BraunsteinPRL1994,Bures1969,Uhlmann1986,HubnerPLA1992} and Hellinger \citep{LuoPRA2004} distances which are referred to as two different extensions from  classical Fisher information.

Parameter encoding can occur in a noisy \cite{SarovarJPA2006,MonrasPRL2007,WatanabePRL2010,LuPRA2010,EscherNP2011,MaPRA2011,ChinPRL2012,BerradaPLA2012,ZhongPRA2013,BerradaPRA2013,OzaydinPLA2014,AlipourPRL2014,BanQIP2015} or noiseless scenario \cite{PangPRA2014,LiuPA2014,LiuCTP2014,LiuSR2015,JingPRA2015}. In the noiseless encoding, which is the purpose  of this work, the parameters are encoded  via a unitary operator $U\left(\boldsymbol{\theta} \right)$  on an  initially $\boldsymbol{\theta}$-independent  probe state $\rho_0$
\begin{equation}\label{RhoURho0}
\rho_{\boldsymbol{\theta}}=U\left(\boldsymbol{\theta} \right)\rho_{0} U^{\dagger}\left(\boldsymbol{\theta} \right),
\end{equation}
where    $\boldsymbol{\theta}=\{\theta_1,\cdots,\theta_n\}$  denotes the set of  parameters to be encoded.
In unitary encoding, the most important ingredients for calculating QFI matrix are  the generators of the parameter translations with respect to each parameter $\theta_i$ of the unitary process $U(\boldsymbol{\theta})$
\begin{eqnarray}\label{RhoEvolution}
\partial_{\theta_i}\rho_{\boldsymbol{\theta}}= -i[K_{\theta_i},\rho_{\boldsymbol{\theta}}].
\end{eqnarray}
These generators  capture  all information of the parametrization process and are defined by \cite{BoixoPRL2007}
\begin{eqnarray}\label{K-theta-i}
{K}_{\theta_i}=i\frac{\partial U(\boldsymbol{\theta})}{\partial{\theta_i}}U^\dagger(\boldsymbol{\theta}),
\end{eqnarray}
or equivalently, up to a unitary transformation in the sense of $M_{\theta_i}=U^\dagger(\boldsymbol{\theta}){K}_{\theta_i}U(\boldsymbol{\theta})$,  can be expressed  as \cite{BoixoPRL2007,TaddeiPRL2013}
\begin{eqnarray}\label{M-theta-i}
M_{\theta_i}=iU^\dagger(\boldsymbol{\theta})\frac{\partial U(\boldsymbol{\theta})}{\partial{\theta_i}}.
\end{eqnarray}
If the unitary process is known, then $K_{\theta_i}$ and $M_{\theta_i}$ can be directly calculated by their  definitions.   Also, when the  estimation parameter $\theta$  is an overall multiplicative factor of the Hamiltonian, the derivative involved in Eqs. \eqref{K-theta-i} and \eqref{M-theta-i} can be calculated straightforwardly. For  estimation of an arbitrary parameter of a $d$-dimensional  Hamiltonian, a  general solution for $K_{\theta}$ is presented in \cite{PangPRA2014}
\begin{eqnarray}
K_{\theta}&=&t\sum_{k=1}^{r}\Tr\{\Gamma_k^\dagger\partial_{\theta}H_{\theta}\}\Gamma_k \\ \nonumber &-&i\sum_{k=r+1}^{d^2}\frac{1-\e^{-i\lambda_kt}}{\lambda_k}\Tr\{\Gamma_k^\dagger\partial_{\theta}H_{\theta}\}\Gamma_k
\end{eqnarray}
where $\lambda_k$ (with $\lambda_k=0$ for $k=1,\cdots,r$ and $\lambda_k\ne 0$ for $k=r+1,\cdots,d^2$) and $\Gamma_k$ are, respectively, eigenvalues and orthonormal eigenvectors of a Hermitian superoperator corresponding to the   Hamiltonian $H_\theta$, obtained from $[H_{\theta},\Gamma]=\lambda\Gamma$.

Moreover, an expanded form for $\mathcal{H}_{\theta_i}=-M_{\theta_i}$ is presented in \cite{LiuSR2015} which requires  calculating an infinite series of
\begin{eqnarray}\label{H}
\mathcal{H}_{\theta_i}=i \sum_{n=0}^{\infty}\mathcal{F}_{n}H_{\boldsymbol{\theta}}^{\times n}(\partial_{\theta_i}H_{\boldsymbol{\theta}}),
\end{eqnarray}
where  $\mathcal{F}_{n}=(it)^{n+1}/(n+1)!$, $H_{\boldsymbol{\theta}}$ is the Hamiltonian of the unitary process $U_{\boldsymbol{\theta}}=\e^{-itH_{\boldsymbol{\theta}}}$, and  $H^{\times n}(\cdot)=[H,\cdots,[H,\cdot]]$.
Utilizing the eigenspectral of $\rho_{0}$
\begin{eqnarray}\label{Rho0}
\rho_{0}=\sum_{a=0}^{s-1}p_{a}\vert\varphi_{a}\rangle\langle\varphi_{a}\vert,
\end{eqnarray}
where $\{p_{a}\}_{a=0}^{s-1}$ and $\{\vert\varphi_{a}\rangle\}_{a=0}^{s-1}$ are the  sets of eigenvalues and eigenvectors  of $\rho_{0}$, respectively, and $s$ is the dimension of the support of $\rho_{0}$, the matrix elements of QFI for  a general unitary transformation $\rho_{\boldsymbol{\theta}}=U\left(\boldsymbol{\theta} \right)\rho_{0} U^{\dagger}\left(\boldsymbol{\theta} \right) $ can be expressed by  \cite{LiuSR2015,LiuCTP2014,LiuPA2014}
\begin{eqnarray}\label{F2-Liu}
F_{\theta_i\theta_j}&=&\sum_{a=0}^{s-1}4p_{a}\textrm{cov}_a(\mathcal{H}_{\theta_i},\mathcal{H}_{\theta_j}) \\ \nonumber
& -&\sum_{a\neq b}\frac{8p_{a}p_{b}}{p_{a}+p_{b}}\textrm{Re}\left\{ \langle\varphi_{a}\vert \mathcal{H}_{\theta_i}\vert\varphi_{b}\rangle\langle\varphi_{b}\vert \mathcal{H}_{\theta_j}\vert\varphi_{a}\rangle\right\},
\end{eqnarray}
where
\begin{eqnarray}\nonumber
\textrm{cov}_a(\mathcal{H}_{\theta_i},\mathcal{H}_{\theta_j})&=&\frac{1}{2}\langle\varphi_{a}\vert \{\mathcal{H}_{\theta_i},\mathcal{H}_{\theta_j}\}\vert\varphi_{a}\rangle \\
&-&\langle\varphi_{a}\vert \mathcal{H}_{\theta_i}\vert\varphi_{a}\rangle\langle\varphi_{a}\vert \mathcal{H}_{\theta_j}\vert\varphi_{a}\rangle
\end{eqnarray}
is the covariance matrix on the  eigenstate $\ket{\varphi_{a}}$ of the initial state  \cite{LiuSR2015,LiuCTP2014,LiuPA2014}.

In this paper, we consider the QFI of a  unitary process  and provide a new representation for QFI of a general $SU(2)$ process. In this representation we associate to each parameter $\theta_i$  a real vectors $\boldsymbol{m}_{\theta_i}\in \mathbb{R}^3$.  The formulation is independent of the  parametrization of the process in a sense that it takes a covariant form for any parametrization  of the process.
We then provide   an explicit  relation for the vectors  $\boldsymbol{m}_{\theta_i}$  for  a general Hamiltonian with arbitrary parametrization.
Furthermore, we present a linear transformation  between two different parameter spaces of a unitary process, enabling  us to interplay between their corresponding  QFI  matrices.
Using this linear transformation, one can go  from either parametrization to  another one, in particular, one can obtain the  QFI matrix of the coset parameters in terms of the one of the angle-axis parameters.

This paper is organized as follows. In section II,  we briefly  review the  QFI and present   a  representation for the QFI matrix of a general unitary process in terms of  the matrices $\boldsymbol{M}_{\theta_i}$.  We then concern ourselves with the particular case of  $SU(2)$ processes  and  introduce  vectors $\boldsymbol{m}_{\theta_i}$, associated with matrices $\boldsymbol{M}_{\theta_i}$, and present a closed relation for QFI matrix in terms of these vectors. An analytical closed relation to evaluate these vectors for general Hamiltonian and arbitrary estimation parameters is also provided in this section. Section III is devoted to present a linear transformation between two different parameter spaces of a unitary process. A way to move from QFI matrix of a unitary process in a given parametrization to the one of the other parametrization is provided in this section. The utility of this transformation is examined by providing an example in qubit systems. The paper is concluded in section IV with a brief discussion.

\section{Quantum Fisher information}
From  various different versions of QFI, the so-called symmetric logarithmic derivative (SLD) Fisher information is the one which has attracted much attention. For a single parameter $\theta$, the SLD Fisher information is defined by \cite{BraunsteinPRL1994,BraunsteinAP1996,PetzJPA2002,PetzBook2008}
\begin{eqnarray}
F_{\theta}=\Tr\left(\rho_{\theta}L_{\theta}^{2}\right),
\label{F.THETA}
\end{eqnarray}
where $\rho_{\theta}$ is the density matrix depending  on $\theta$,  and $L_{\theta}$ is the SLD operator  determined by the equation
\begin{eqnarray}\label{RHO.L}
\partial_\theta\rho_{\theta}=\frac{1}{2}
&\left\{ L_{\theta},\rho_{\theta}\right\},
\end{eqnarray}
where $\left\lbrace ,\right\rbrace $ denotes anticommutator. For a multiparameter scenario    $\boldsymbol{\theta}=\{\theta_1,\cdots,\theta_n\}$, the quantum Fisher information  matrix is defined by
\begin{eqnarray}\label{Fij}
F_{\theta_i\theta_j}=\frac{1}{2}\Tr\left( \rho\left\lbrace L_{\theta_i},L_{\theta_j} \right\rbrace \right),
\end{eqnarray}
where $L_{\theta_i}$ is the SLD operator for the parameter $\theta_i$, given by
\begin{eqnarray}\label{RhoLi}
\partial_{\theta_i}\rho_{\boldsymbol{\theta}}=\frac{1}{2}\left\{ L_{\theta_i},\rho_{\boldsymbol{\theta}}\right\},
\end{eqnarray}
and $L_{\theta_j}$  is   defined similarly.

Using the eigenspectral of $\rho_{0}$ given in Eq. \eqref{Rho0},  one can write the eigenspectral  of $\rho_{\boldsymbol{\theta}}$ as
\begin{eqnarray}\label{RhoTheta}
\rho_{\boldsymbol{\theta}}=\sum_{a=0}^{s-1}p_{a}\vert\tilde{\varphi}_{a}\rangle\langle\tilde{\varphi}_{a}\vert,
\end{eqnarray}
where $\ket{\tilde{\varphi}_{a}}=U\left(\boldsymbol{\theta} \right)\ket{\varphi_{a}}$ denotes  eigenvectors  of $\rho_{\boldsymbol{\theta}}$. In this basis  Eqs. \eqref{RhoEvolution} and \eqref{RhoLi}  read, respectively, as (for $i=1,\cdots,n$)
\begin{eqnarray}\nonumber
\left( \partial_{\theta_i}\rho_{\boldsymbol{\theta}}\right)_{\tilde{a}\tilde{b}}&=&i(p_{a}-p_{b})\left( K_{\theta_i}\right)_{\tilde{a}\tilde{b}} \\ \nonumber
&=&\frac{1}{2}(p_{a}+p_{b})\left( L_{\theta_i}\right)_{\tilde{a}\tilde{b}},
\label{EM.L}
\end{eqnarray}
where we have defined  $(K_{\theta_i})_{\tilde{a}\tilde{b}}=\langle\tilde{\varphi}_{a}\vert K_{\theta_i} \vert\tilde{\varphi}_{b}\rangle$ and $(L_{\theta_i})_{\tilde{a}\tilde{b}}=\langle\tilde{\varphi}_{a}\vert L_{\theta_i} \vert\tilde{\varphi}_{b}\rangle$.
Using this,  one can find the matrix elements of the SLD operators in the $\boldsymbol{\theta}$-parametrization in terms of the matrix elements of the corresponding matrices $K_{\theta_i}$  as
\begin{eqnarray}\label{Lthetai}
(L_{\theta_i})_{\tilde{a}\tilde{b}}=2i\frac{\left(p_{a}-p_{b}\right)}{p_{a}+p_{b}}(K _{\theta_i})_{\tilde{a}\tilde{b}}.
\end{eqnarray}
This  can be used in  Eq. \eqref{Fij} to find matrix  elements of the QFI matrix in the $\boldsymbol{\theta}$-representation as
\begin{eqnarray} \nonumber
F_{\theta_i\theta_j}&=&\sum_{a}\sum_{b}2\frac{(p_{a}-p_{b})^2}{p_{a}+p_{b}}(K_{\theta_i})_{\tilde{a}\tilde{b}}(K_{\theta_j})_{\tilde{b}\tilde{a}} \\ \label{FisherMatrix}
&=&\sum_{a}\sum_{b}2\frac{(p_{a}-p_{b})^2}{p_{a}+p_{b}}(M_{\theta_i})_{{a}{b}}(M_{\theta_j})_{{b}{a}},
\end{eqnarray}
where $({M} _{\theta_i})_{ab}=\bra{\varphi_a}{M}_{\theta_i}\ket{\varphi_b}$.
Equation \eqref{FisherMatrix} provides a
 relation for the QFI matrix of an arbitrary  unitary process $U(\boldsymbol{\theta})$, and is  equivalent to the one presented by Eq. \eqref{F2-Liu} \cite{LiuSR2015}. Accordingly, the QFI matrix can be calculated  provided that we could calculate the infinitesimal generators   $K _{\theta_i}$ (or   $M _{\theta_i}$)
associated to each parameter $\theta_i$ ($i=1,\cdots,n$).
Instead of using matrix representation of operators, a useful technique is to utilize the Bloch vector representation. This method has been used recently for the SLD operator   to derive an explicit expression for the Holevo bound for estimating two-parameter family of qubit states \cite{SuzukiIJQI2015,SuzukiJMP2016}.   In the next section we concern our attention to the $SU(2)$ processes  and by using  the Bloch vector  representation for  matrices $M_{\theta_i}$, Eq. \eqref{M-theta-i},
 a closed relation for  the QFI matrix   of arbitrary parameters of a general Hamiltonian is provided.

\subsection{SU(2) processes}
For the  simplest case of $SU(2)$ processes we will provide a closed relation for Eq. \eqref{FisherMatrix}  in terms of the Bloch vector representation of the $M$-matrices. To do so, first suppose  that the initial state $\rho_0$ is diagonal in the computational  basis  $\{\ket{0},\ket{1}\}$. In this case  Eq. \eqref{FisherMatrix} reduces to
\begin{eqnarray}\nonumber
F_{\theta_i\theta_j}&=&2(p_{0}-p_{1})^2\left((M_{\theta_i})_{01}(M_{\theta_j})_{10}+(M_{\theta_i})_{10}(M_{\theta_j})_{01}\right) \\
&=&2(p_{0}-p_{1})^2\left(\Tr{[M_{\theta_i}M_{\theta_j}]}-2(M_{\theta_i})_{00}(M_{\theta_j})_{00}
\right),
\end{eqnarray}
where in the last line we have used the fact that $M$-matrices are traceless. Now, to each Hermitian traceless $2\times 2$ matrix $M_{\theta_i}$, one can associate
a real vector $\boldsymbol{m}_{\theta_i}\in \mathbb{R}^3$ by
$M_{\theta_i}=\boldsymbol{\sigma}\cdot\boldsymbol{m}_{\theta_i}$. In this representation we have $(M_{\theta_i})_{00}=(\boldsymbol{\sigma}\cdot\boldsymbol{m}_{\theta_i})_{00}=\hat{\boldsymbol{z}}\cdot\boldsymbol{m}_{\theta_i}$  and
$M_{\theta_i}M_{\theta_j}=(\boldsymbol{\sigma}\cdot\boldsymbol{m}_{\theta_i})(\boldsymbol{\sigma}\cdot\boldsymbol{m}_{\theta_j})
=(\boldsymbol{m}_{\theta_i}\cdot \boldsymbol{m}_{\theta_j})\Id_2+i\boldsymbol{\sigma}\cdot (\boldsymbol{m}_{\theta_i}\times \boldsymbol{m}_{\theta_j})$. We find
\begin{eqnarray}\label{FisherQubit-z}
F_{\theta_i\theta_j}=4(p_{0}-p_{1})^2\left[\boldsymbol{m}_{\theta_i}\cdot \boldsymbol{m}_{\theta_j}-(\hat{\boldsymbol{z}}\cdot\boldsymbol{m}_{\theta_i})(\hat{\boldsymbol{z}}\cdot\boldsymbol{m}_{\theta_j})
\right].
\end{eqnarray}
In general, however,  we are interested in the QFI of the unitary process $U(\boldsymbol{\theta})$
 starting from an arbitrary initial state $\rho_0$ with associated orthonormal eigenbasis $\{\ket{\varphi_0},\ket{\varphi_1}\}$. To do this
we define  $\ket{\varphi_a}=\Omega(\theta,\phi)\ket{a}$, for $a=0,1$, with
\begin{equation}\label{OmegaThetaPhi}
\Omega(\theta,\phi)=\left(\begin{array}{cc}\cos{\frac{\theta}{2}} & -\e^{-i\phi}\sin{\frac{\theta}{2}} \\ \e^{i\phi}\sin{\frac{\theta}{2}} & \cos{\frac{\theta}{2}}\end{array}\right).
\end{equation}
One can easily show that $\ket{\varphi_0}$ and $\ket{\varphi_1}$ are eigenvectors of $\boldsymbol{\sigma}\cdot \hat{\boldsymbol{n}}$ corresponding to the eigenvalues $+1$ and $-1$,
respectively, where $\hat{\boldsymbol{n}}=(\sin{\theta}\cos{\phi}, \sin{\theta}\sin{\phi}, \cos{\theta})^{\T}$.
Starting from this initial state transforms  the $M$-matrices as $M_{\theta_i} \rightarrow M_{\theta_i}=\Omega^\dagger(\theta,\phi) M_{\theta_i} \Omega(\theta,\phi)$.
Associated to this unitary transformation  the $\boldsymbol{m}$-vectors rotate  as $\boldsymbol{m}_{\theta_i} \rightarrow O^\T\boldsymbol{m}_{\theta_i}$, where
 the orthogonal matrix $O$ is defined  by $O_{ij}=\frac{1}{2}\Tr{[\sigma_i \Omega \sigma_j \Omega^\dagger]}$.
Obviously, $\boldsymbol{m}_{\theta_i}\cdot \boldsymbol{m}_{\theta_j}$ remains invariant under such transformation
and  that $\hat{\boldsymbol{z}}\cdot O^\T\boldsymbol{m}_{\theta_i}=O\hat{\boldsymbol{z}}\cdot \boldsymbol{m}_{\theta_i}$. Moreover, simple  calculation shows that
for the unitary transformation \eqref{OmegaThetaPhi},  $O\hat{\boldsymbol{z}}$ is nothing but the unit vector $\hat{\boldsymbol{n}}$ defined above. We therefore  arrive at  the following proposition  for the QFI matrix of the unitary process $U(\boldsymbol{\theta})\in SU(2)$.
\begin{proposition}
To each parameter $\theta_i$ of the unitary process $U(\boldsymbol{\theta})$ one can associate a unique  vector $\boldsymbol{m}_{\theta_i}$  defined by $[\boldsymbol{m}_{\theta_i}]_k=\frac{1}{2}\Tr\{\sigma_kM_{\theta_i}\}$, where $M_{\theta_i}$ is given by  Eq. \eqref{M-theta-i}. Using this, the QFI matrix takes the following form
\begin{eqnarray} \label{FisherQubit-n}
F_{\theta_i\theta_j}
&=&4(p_{0}-p_{1})^2\left[\boldsymbol{m}_{\theta_i}^\T\Lambda_{\hat{\boldsymbol{n}}} \boldsymbol{m}_{\theta_j}\right] \\ \nonumber
&=& 4(p_{0}-p_{1})^2\left[\boldsymbol{m}_{\theta_i}\cdot \boldsymbol{m}_{\theta_j}
-(\hat{\boldsymbol{n}}\cdot\boldsymbol{m}_{\theta_i})(\hat{\boldsymbol{n}}\cdot\boldsymbol{m}_{\theta_j})
\right]
\end{eqnarray}
where $\Lambda_{\hat{\boldsymbol{n}}}=\Id_3-\hat{\boldsymbol{n}}\hat{\boldsymbol{n}}^\T$ is a two-dimensional projection operator orthogonal to  $\hat{\boldsymbol{n}}$.
\end{proposition}
This  simple form shows that the QFI matrix of a unitary process is  composed of two independent contributions;
first,  each parameter $\theta_i$ of   the unitary process $U(\boldsymbol{\theta})$ is contributed in the Fisher information via the    vector $\boldsymbol{m}_{\theta_i}\in \mathbb{R}^3$,
and second,  the role of the initial state $\rho_0$ is played  by the  Bloch vector $\hat{\boldsymbol{n}}$ and the eigenvalues $p_0,p_1$.
However, looking at  Eq. \eqref{FisherQubit-n}  shows that  although $\boldsymbol{m}_{\theta_i}$ are vectors in $\mathbb{R}^3$, their role in the QFI matrix  is played  effectively in a two dimensional subspace perpendicular to  $\hat{\boldsymbol{n}}$.
To see this note that $\boldsymbol{m}_{\theta_i}^\T\Lambda_{\hat{\boldsymbol{n}}} \boldsymbol{m}_{\theta_j}
=(\Lambda_{\hat{\boldsymbol{n}}}\boldsymbol{m}_{\theta_i})^\T(\Lambda_{\hat{\boldsymbol{n}}} \boldsymbol{m}_{\theta_j})$
 and  $\Lambda_{\hat{\boldsymbol{n}}}\boldsymbol{m}_{\theta_i}\in \mathrm{Range}\{\Lambda_{\hat{\boldsymbol{n}}}\}$. Accordingly, initial states with different Bloch vectors result in different  subspaces, hence  different QFI matrix, in general. It turns out that   the QFI matrix  is invariant under orthogonal transformation on the vectors $\boldsymbol{m}_{\theta_i}$, i.e. $\boldsymbol{m}_{\theta_i} \longrightarrow R\boldsymbol{m}_{\theta_i}$, provided that the Bloch vector of the initial state is changed as $\hat{\boldsymbol{n}} \longrightarrow R^\T \hat{\boldsymbol{n}}$. In view of this, the maximum of QFI matrix over all initial states, if exists,  is invariant under orthogonal transformation performed on $\boldsymbol{m}_{\theta_i}$. For instance, for a single parameter $\theta$, Eq. \eqref{FisherQubit-n} reduces to $F_{\theta}=4(p_{0}-p_{1})^2\left[|\boldsymbol{m}_{\theta}|^2-(\hat{\boldsymbol{n}}\cdot\boldsymbol{m}_{\theta})^2\right]$, implies that the QFI  attains its maximum value $F_{\theta}^{\max}=4|\boldsymbol{m}_{\theta}|^2$, gained by  any  initial pure state $\ket{\boldsymbol{\sigma}\cdot \hat{\boldsymbol{n}}}$ with $\hat{\boldsymbol{n}}$ lying in the plane perpendicular to $\boldsymbol{m}_{\theta}$.

Another consequence of  Eq. \eqref{FisherQubit-n} is that the lack of independency of the vectors $\boldsymbol{m}_{\theta_i}$ leads to  vanishing determinant of the QFI matrix, meaning that the variances of the set of parameters cannot be estimated simultaneously through the Cramer-Rao  bound.
To make an interpretation of this, suppose   $\boldsymbol{m}_{\theta_k}=\sum_{l\ne k}c_l\boldsymbol{m}_{\theta_l}$ for some nonzero real numbers $c_l$. In this case  we find $M_{\theta_k}=\sum_{l\ne k}c_lM_{\theta_l}$, results in $\frac{\partial U(\boldsymbol{\theta})}{\partial{\theta_k}}=\sum_{l\ne k}c_l\frac{\partial U(\boldsymbol{\theta})}{\partial{\theta_l}}$ from Eq. \eqref{M-theta-i}. Defining $U(\boldsymbol{\theta})=\e^{-iHt}$ for  Hamiltonian $H$ and invoking Eq. \eqref{Wilcox}, we find  $\frac{\partial H}{\partial{\theta_k}}=\sum_{l\ne k}c_l\frac{\partial H}{\partial{\theta_l}}$. The converse is also true meaning that any relation between derivatives of the Hamiltonian  leads to the same relation between the corresponding $\boldsymbol{m}$-vectors, as such,  for any initial probe state $\rho_0$ the QFI matrix becomes singular. What is noteworthy here is that the  lack of independency of $\boldsymbol{m}$-vectors is not necessary to get singular QFI matrix, as Eq. \eqref{FisherQubit-n} could lead to   a singular QFI matrix even for linearly independent $\boldsymbol{m}$-vectors. To see this consider, for example,  two arbitrary parameters $\theta_1$ and $\theta_2$   associated with two linearly independent vectors   $\boldsymbol{m}_{\theta_1}$ and $\boldsymbol{m}_{\theta_2}$. One can see that for any  Bloch vector of the initial state lying  in the plane of $\boldsymbol{m}_{\theta_1}$ and $\boldsymbol{m}_{\theta_2}$, i.e. for any $\hat{\boldsymbol{n}}=a_1\boldsymbol{m}_{\theta_1}+a_2\boldsymbol{m}_{\theta_2}$ with arbitrary  real numbers $a_1$ and  $a_2$ such that  $a_1^2|\boldsymbol{m}_{\theta_1}|^2+a_2^2|\boldsymbol{m}_{\theta_2}|^2+2a_1a_2\boldsymbol{m}_{\theta_1}\cdot\boldsymbol{m}_{\theta_2}=1$,  the QFI matrix obtained from Eq. \eqref{FisherQubit-n} has a vanishing determinant.

Moreover,
for the eigendecomposition of $\rho_{\boldsymbol{\theta}}$, given by Eq. \eqref{RhoTheta}, one can see that $F(\rho_{\boldsymbol{\theta}})=(p_0-p_1)^2F(\tilde{\varphi}_a)$ for $a=0,1$, as such $F(\rho_{\boldsymbol{\theta}})/\sum_{a=0}^{1}p_aF(\tilde{\varphi}_a)=(p_0-p_1)^2\le 1$, implies convexity of the QFI in this case.

Having Eq. \eqref{FisherQubit-n} as  a relation  for QFI matrix in terms of the vectors $\boldsymbol{m}_{\theta_i}$, it is now the time to present a relation  to calculate the required vectors $\boldsymbol{m}_{\theta_i}$ for a general Hamiltonian. The following proposition  provides an explicit representation for vectors $\boldsymbol{m}_{\theta_i}$ of a general qubit Hamiltonian.
\begin{proposition}
For a unitary process generated by the Hamiltonian \cite{JingPRA2015}
\begin{equation}\label{Hamiltonian-1}
H=\boldsymbol{\alpha}\cdot \boldsymbol{\sigma},
\end{equation}
the associated $\boldsymbol{m}$-vectors are given by the following relation
\begin{eqnarray}\label{m-theta-i-1}
\boldsymbol{m}_{\theta_i}=\frac{\partial {|\boldsymbol{\alpha}|}}{\partial{\theta_i}}t\;\hat{\boldsymbol{\alpha}}&+&\sin{(|\boldsymbol{\alpha}| t)}\cos{(|\boldsymbol{\alpha}| t)}\frac{\partial {\hat{\boldsymbol{\alpha}}}}{\partial{\theta_i}} \\ \nonumber
&-& \sin^2{(|\boldsymbol{\alpha}| t)}\left(\hat{\boldsymbol{\alpha}}\times \frac{\partial {\hat{\boldsymbol{\alpha}}}}{\partial{\theta_i}}\right),
\end{eqnarray}
where we have defined the unit vector $\hat{\boldsymbol{\alpha}}=\boldsymbol{\alpha}/|\boldsymbol{\alpha}|$.
\end{proposition}
Before we proceed further to provide  a proof for the above equation, we have to stress here that a similar relation, however with different derivation, is provided in Ref. \cite{JingPRA2015}.
\begin{proof}
Using the equation \cite{WilcoxJMP1967}
\begin{equation}\label{Wilcox}
\frac{\partial }{\partial \theta_i}\e^{-iHt}=\int_{0}^{1}\e^{-isHt}\left(-it\frac{\partial }{\partial \theta_i}H\right)\e^{-i(1-s)Ht} \dif s,
\end{equation}
in the definition of $[\boldsymbol{m}_{\theta_i}]_k$, we get
\begin{eqnarray}\nonumber
[\boldsymbol{m}_{\theta_i}]_k&=&\frac{i}{2}\Tr\left\{\sigma_k U^\dagger(\boldsymbol{\theta})\frac{\partial U(\boldsymbol{\theta})}{\partial{\theta_i}}\right\} \\ \nonumber
&=&\frac{t}{2}\int_{0}^{1}\dif s\;\Tr\left\{V(s)\sigma_k V^\dagger(s)\frac{\partial H}{\partial{\theta_i}}\right\} \\
&=&t\sum_{l=1}^{3}\frac{\partial \alpha_l}{\partial{\theta_i}}\int_{0}^{1}\dif s\;O_{lk}(s),
\label{m-theta-i-k}
\end{eqnarray}
where we have defined $V(s)=\e^{-i(1-s)Ht}$, and  $O_{lk}(s)=\frac{1}{2}\Tr\left\{V(s)\sigma_k V^\dagger(s)\sigma_l\right\}$ is the orthogonal matrix corresponding to the unitary matrix $V(s)$. Now,  for $H=\boldsymbol{\alpha}\cdot \boldsymbol{\sigma}$, one can use $V(s)=\Id_2 \cos[\tau]-i\hat{\boldsymbol{\alpha}}\cdot \boldsymbol{\sigma}\sin[\tau]$ with $\tau=(1-s)|\boldsymbol{\alpha}|t$, so that
\begin{eqnarray}\label{O-lk}
O_{lk}(s)=\cos[2\tau]\delta_{kl}-\sin[2\tau]\varepsilon_{ktl}\hat{\alpha}_{t}+2\sin^2[\tau]\hat{\alpha}_{k}\hat{\alpha}_{l},\quad
\end{eqnarray}
where $\varepsilon_{ktl}$ is the so-called Levi-Civita symbol and  summation over repeated indices is understood. Using this in Eq. \eqref{m-theta-i-k} and after calculating  the integrals, we get
\begin{eqnarray}\nonumber
\boldsymbol{m}_{\theta_i}&=&\frac{1}{2|\boldsymbol{\alpha}|}\left\{\sin(2|\boldsymbol{\alpha}| t)\frac{\partial {\boldsymbol{\alpha}}}{\partial{\theta_i}}-(1-\cos(2|\boldsymbol{\alpha}| t))\left(\hat{\boldsymbol{\alpha}}\times \frac{\partial {\boldsymbol{\alpha}}}{\partial{\theta_i}}\right)\right. \\
&+& \left. \left[(2|\boldsymbol{\alpha}| t-\sin(2|\boldsymbol{\alpha}|t))\left(\hat{\boldsymbol{\alpha}} \cdot \frac{\partial {{\boldsymbol{\alpha}}}}{\partial{\theta_i}}\right)\right]\hat{\boldsymbol{\alpha}}\right\}.
\end{eqnarray}
Finally, using $\boldsymbol{\alpha}=|\boldsymbol{\alpha}|\hat{\boldsymbol{\alpha}}$ and noting that
$\hat{\boldsymbol{\alpha}}\cdot \frac{\partial {\hat{\boldsymbol{\alpha}}}}{\partial{\theta_i}}=0$,
 we find Eq.  \eqref{m-theta-i-1}.
\end{proof}
Note that in  Eq. \eqref{m-theta-i-1}   we have not fixed  the parameters under estimation, in a sense that both amplitude and direction of the vector $\boldsymbol{\alpha}$ can be depend on each of the parameters $\theta_i$. Moreover,  the $\boldsymbol{m}$-vectors  are generally not orthogonal nor normalized. In the following we will consider the so-called angle-axis parametrization of the $SU(2)$ group and show that for such a set of parameters the associated $\boldsymbol{m}$-vectors are orthogonal.

{\textit Example.---}Consider  a system described by  the Hamiltonian \eqref{Hamiltonian-1}, with   $\boldsymbol{\alpha}$ described by the following relation  \citep{JingPRA2015}
\begin{eqnarray}\label{alpha-rtp}
\boldsymbol{\alpha}=r\hat{\boldsymbol{\alpha}}, \quad   \hat{\boldsymbol{\alpha}}=(\sin\vartheta\cos\varphi, \sin\vartheta\sin\varphi, \cos\vartheta)^\T.
\end{eqnarray}
The unitary evolution generated by  this Hamiltonian is given by $U(r,\vartheta,\varphi)=\e^{-iH t}$.  Taking  $\theta_1=r$,  $\theta_2=\vartheta$, and  $\theta_3=\varphi$ as the parameters under estimation,   one can easily find from Eq. \eqref{m-theta-i-1}
\begin{eqnarray}\label{m-canonic-1}
\boldsymbol{m}_{r}&=&t\hat{\boldsymbol{\alpha}}_{0}, \\ \label{m-canonic-2}
\boldsymbol{m}_{\vartheta}&=&\sin{rt}\left[\cos{rt}\; \hat{\boldsymbol{\alpha}}_{1}-\sin{rt}\;\hat{\boldsymbol{\alpha}}_{2}\right],\\ \label{m-canonic-3}
\boldsymbol{m}_{\varphi}&=&\sin{\vartheta}\sin{rt}\left[\sin{rt}\;\hat{\boldsymbol{\alpha}}_{1}+\cos{rt}\;\hat{\boldsymbol{\alpha}}_{2}\right],
\end{eqnarray}
where
\begin{equation}
\hat{\boldsymbol{\alpha}}_{0}=\hat{\boldsymbol{\alpha}},\qquad
\hat{\boldsymbol{\alpha}}_{1}=\frac{\partial {\hat{\boldsymbol{\alpha}}}}{\partial{\vartheta}}, \qquad
\hat{\boldsymbol{\alpha}}_{2}=\frac{1}{\sin{\vartheta}}\frac{\partial {\hat{\boldsymbol{\alpha}}}}{\partial{\varphi}},
\end{equation}
form an orthonormal basis. Clearly, such a set of $\boldsymbol{m}$-vectors is orthogonal and can  be written as
\begin{eqnarray}\label{m-R}
\boldsymbol{m}_r=\mathcal{R} \boldsymbol{\mathfrak{m}}_r, \quad
\boldsymbol{m}_\vartheta= \mathcal{R} \boldsymbol{\mathfrak{m}}_\vartheta, \quad
\boldsymbol{m}_\varphi= \mathcal{R} \boldsymbol{\mathfrak{m}}_\varphi,
\end{eqnarray}
with
\begin{eqnarray}
\boldsymbol{\mathfrak{m}}_r&=& t\left(0 \; , \;  0 \; , \;  1 \right)^\T, \\
\boldsymbol{\mathfrak{m}}_\vartheta&=& \sin{rt}\left(\cos{rt} \; , \; -\sin{rt} \; , \; 0 \;  \right)^\T, \\
\boldsymbol{\mathfrak{m}}_\varphi&=& \sin{\vartheta}\sin{rt}\left( \sin{rt} \; , \; \cos{rt} \; , \; 0 \right)^\T,
\end{eqnarray}
Above, $\mathcal{R}=\mathcal{R}_z(\varphi)\mathcal{R}_y(\vartheta)$ where $\mathcal{R}_z(\varphi)$  and  $\mathcal{R}_y(\vartheta)$ denote  rotations about $z$ and $y$ axes, respectively
\begin{eqnarray}
\mathcal{R}_z(\varphi)&=&\begin{pmatrix}
\cos{\varphi} & -\sin{\varphi} & 0 \\
\sin{\varphi} & \cos{\varphi} & 0 \\
0 & 0 & 1
\end{pmatrix}, \\
\mathcal{R}_y(\vartheta)&=&\begin{pmatrix}
\cos{\vartheta} & 0 & \sin{\vartheta} \\
0 & 1 & 0 \\
-\sin{\vartheta} & 0 & \cos{\vartheta} &
\end{pmatrix}.
\end{eqnarray}
Note that  vectors  $\boldsymbol{\mathfrak{m}}_r$, $\boldsymbol{\mathfrak{m}}_\vartheta$, and $\boldsymbol{\mathfrak{m}}_\varphi$ are independent of the azimuthal angle $\varphi$.   With these $\boldsymbol{m}$-vectors in hand,  one can easily use Eq. \eqref{FisherQubit-n} to calculate QFI of the parameters  $r$, $\vartheta$, and $\varphi$  for an arbitrary initial state.
We find
\begin{eqnarray}\nonumber
F_{\theta_i\theta_j}=4(p_{0}-p_{1})^2\left[\boldsymbol{\mathfrak{m}}_{\theta_i}^\T\Lambda_{\hat{\boldsymbol{n}}^\prime} \boldsymbol{\mathfrak{m}}_{\theta_j}\right],
\end{eqnarray}
where $\Lambda_{\hat{\boldsymbol{n}}^\prime}=\Id_3-\hat{\boldsymbol{n}}^\prime {\hat{\boldsymbol{n}}^{\prime^\T}}$ with $\hat{\boldsymbol{n}}^\prime=\mathcal{R}^\T\hat{\boldsymbol{n}}$.

\begin{figure}[t!]
\centering
\includegraphics[scale=0.8]{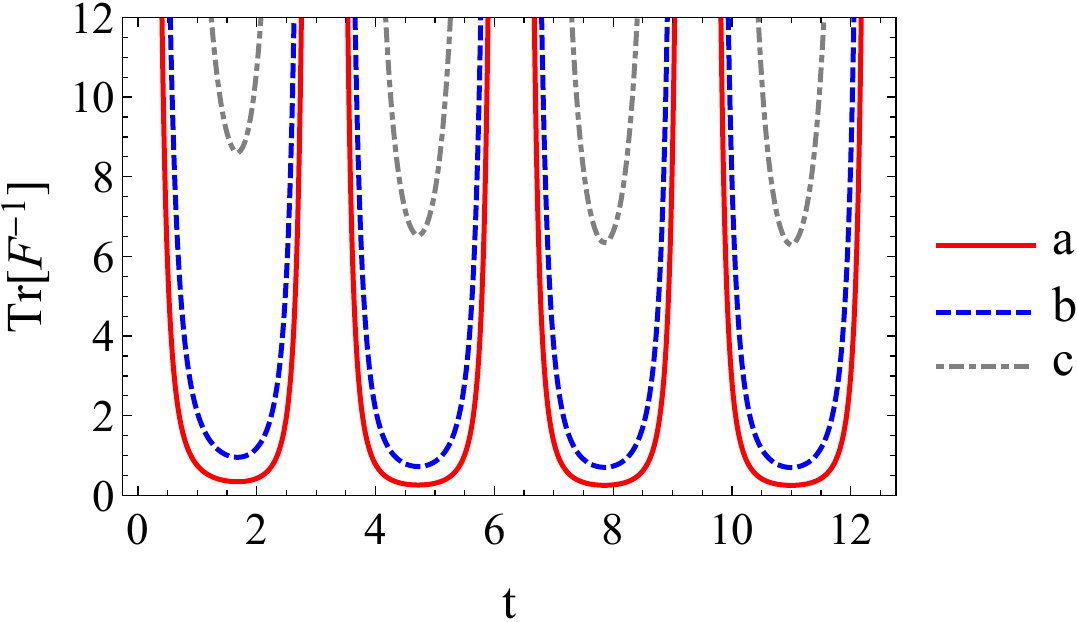}
\caption{(Color online) Total precision limit, given by Eq. \eqref{TPL}, as a function of time for $B=1$,  $\theta=\pi/2$, and different values of $p_0$ ($p_1=1-p_0$). (a)  $p_{0}=1$ (solid-red line). (b)  $p_{0}=0.8$ (dashed-blue line).   (c)  $p_{0}=0.6$ (dot-dashed-gray line). }
\label{Fig-TPL}
\end{figure}

As a particular case consider    a spin-half system in a magnetic field $B$ described by the Hamiltonian
\begin{eqnarray}
H_{\vartheta}=B\left( \sin{\vartheta}\sigma_{1}+\cos{\vartheta}\sigma_{3}\right).
\end{eqnarray}
This Hamiltonian can be obtained from  Eqs. \eqref{Hamiltonian-1} and \eqref{alpha-rtp} by setting  $r\rightarrow B$, $\vartheta \rightarrow \vartheta$ and $\varphi\rightarrow 0$. Suppose that the magnetic field $B$ is known and $\vartheta$ is the parameter under estimation. In this case we find $\boldsymbol{\mathfrak{m}}_\vartheta=\sin{Bt} \left(\cos{Bt} \; , \; -\sin{Bt} \; , \; 0 \;  \right)^\T$, so that
\begin{eqnarray}\label{m-vartheta}
\boldsymbol{m}_\vartheta&=&\sin{Bt}\left(\cos\vartheta\cos{Bt} \;,\; -\sin{Bt}\; ,\; -\sin\vartheta\cos{Bt}\right)^{\T}.
\end{eqnarray}
Using this in Eq. \eqref{FisherQubit-n} one can easily find the QFI.
In this case  the maximum Fisher information leads \cite{PangPRA2014}
\begin{eqnarray}
F_\vartheta^{\max }=4|\boldsymbol{m}_\vartheta|^2= 4\sin^{2}{(Bt)},
\end{eqnarray}
which  happens for any initial pure state with Bloch vector $\hat{\boldsymbol{n}}$ perpendicular to $\boldsymbol{m}_\vartheta$.
If  both $B$ and $\vartheta$ are parameters under estimation,  the $\boldsymbol{m}_\vartheta$ is given by the same Eq. \eqref{m-vartheta}, and $\boldsymbol{m}_B$ is defined by
\begin{equation}\label{m-B}
\boldsymbol{m}_B=t\left(\sin{\vartheta} \; , \;  0 \; , \; \cos{\vartheta}\right)^{\T}.
\end{equation}
For instance, if  the initial state is taken in the spin $\hat{\boldsymbol{z}}$-direction, one can find the QFI matrix as
\begin{eqnarray}
F&=&4(p_{0}-p_{1})^2 \\ \nonumber
& \times & \begin{pmatrix}
 \sin^{2}(Bt)\left(1-\sin^{2}{\vartheta}\cos^{2}(Bt)\right) &\frac{t}{4}\sin(2\vartheta)\sin(2Bt)\\
\frac{t}{4}\sin(2\vartheta)\sin(2Bt) & t^2\sin^{2}{\vartheta}
\end{pmatrix}.
\end{eqnarray}
In this case, the ultimate precision limit is given by the trace of inverse of the QFI matrix, i.e.
\begin{eqnarray}\label{TPL}
\Tr{F^{-1}}=\frac{1}{4(p_{0}-p_{1})^2}\left(
 \frac{1}{\sin^{4}(Bt)}+\frac{1-\sin^{2}{\vartheta}\cos^{2}(Bt)}{t^2\sin^2\vartheta\sin^2(Bt)}\right).
\end{eqnarray}
Figure \ref{Fig-TPL} shows the above limit in terms of time for $B=1$,  $\theta=\pi/2$, and different values of $p_0$. As can be seen from this figure the best  achievable precision happens when the initial probe state is pure.

\section{QFI of a unitary process with two different parametrizations}
Let us consider a unitary transformation parameterized in terms of two different classes of parameters $\boldsymbol{\alpha}=\{\alpha_k\}$ and $\boldsymbol{\beta}=\{\beta_{l}\}$, i.e. $U(\boldsymbol{\alpha})=U(\boldsymbol{\beta})$. Now the question is that if we start with the same initial state $\rho_0$ and encode  these parameters on the state as
\begin{eqnarray}
\rho_{\boldsymbol{\alpha}}&=&U(\boldsymbol{\alpha})\rho_{0}U^{\dagger}(\boldsymbol{\alpha}) \\ \nonumber
&=&U(\boldsymbol{\beta})\rho_{0}U^{\dagger}(\boldsymbol{\beta})=\rho_{\boldsymbol{\beta}},
\end{eqnarray}
what is the relation between the QFI matrices of these two parametrizations?

To address this question we should first find a relation between $M$-matrices with respect to these classes of parameters. To do so, we write Eq. \eqref{M-theta-i}  for $\alpha_k$ and $\beta_l$ as ${M}_{\alpha_k}=iU^\dagger\frac{\partial U}{\partial{\alpha_k}}$ and   ${M}_{\beta_l}=iU^\dagger\frac{\partial U}{\partial{\beta_l}}$, respectively.  By using $\frac{\partial U}{\partial\beta_{l}}=\sum_{k}\frac{\partial U}{\partial\alpha_{k}}\frac{\partial\alpha_{k}}{\partial\beta_{l}}$, we get
\begin{eqnarray}\label{MbetaMalpha}
{M}_{\beta_l}=i\sum_{k}U^\dagger\frac{\partial U}{\partial\alpha_{k}}\frac{\partial\alpha_{k}}{\partial\beta_{l}} =\sum_{k}M_{\alpha_k} S_{\alpha_{k},\beta_{l}},
\end{eqnarray}
where we have defined the transfer matrix $S$ with matrix elements $S_{\alpha_{k},\beta_{l}}=\frac{\partial\alpha_{k}}{\partial\beta_{l}}$. Similarly, if we write  Eq. \eqref{Lthetai} for $\alpha_k$ and $\beta_l$ and by using Eq. \eqref{MbetaMalpha} we find a relation between SLD matrices with respect to these classes of parameters
\begin{eqnarray}
\left( L_{\beta_l}\right)_{\tilde{a}\tilde{b}}=\sum_{k}\left( L_{\alpha_k}\right)_{\tilde{a}\tilde{b}}S_{\alpha_{k},\beta_{l}}.
\end{eqnarray}
Finally the relation between various parametrizations of the QFI can be expressed by
\begin{eqnarray}
F_{{\beta_{l}}{\beta_{l^\prime}}}=\sum_{k}\sum_{k^\prime}S_{\alpha_{k},\beta_{l}}F_{{\alpha_{k}}{\alpha_{k^\prime}}}
S_{\alpha_{k^\prime},\beta_{l^\prime}},
\label{FbetaFalpha1}
\end{eqnarray}
which  can be written in a more compact form as
\begin{equation}\label{FbetaFalpha2}
F^{\{\beta_{l}\}}=S^\T F^{\{\alpha_{k}\}}S.
\end{equation}
Very recently a similar relation is presented in Ref. \cite{GoldbergPRA2018}, however it is for the special case of $SU(2)$ processes with the aim of calculating the QFI of  arbitrary parameters of $SU(2)$ using the one of the Euler angles. Regarding this, Eq. \eqref{FbetaFalpha1} is general in a sense that  it enables one to obtain the QFI matrix  of an arbitrary  unitary process for a given set of estimation parameters  from the  one of the other set of parameters, with no restriction on the number of the initial and final   estimation parameters.

For the simplest case of $SU(2)$ processes, the relation between  QFI of different parametrizations can be expressed in terms of a relation between $\boldsymbol{m}$-vectors of the corresponding parameters.   Actually,  if $\boldsymbol{m}_{\alpha_{k}}$  and $\boldsymbol{m}_{\beta_{l}}$ denote the $\boldsymbol{m}$-vectors of a unitary process
in the $\{\alpha_k\}$ and $\{\beta_l\}$  parametrizations, respectively, we find from Eq. \eqref{MbetaMalpha}
\begin{eqnarray}\label{m-beta-m-alpha}
{\boldsymbol{m}}_{\beta_l}=\sum_{k}\boldsymbol{m}_{\alpha_k} S_{\alpha_{k},\beta_{l}}.
\end{eqnarray}
In view of this,   both $F_{\alpha_k\alpha_{k^\prime}}$ and $F_{\beta_l\beta_{l^\prime}}$ are given by the same relation \eqref{FisherQubit-n} with their own $\boldsymbol{m}$-vectors   replaced by $\boldsymbol{m}_{\theta_{i}}$.
In order to show how the above algorithm works, in the example below  we will obtain the QFI of a unitary process in the coset representation from the one in the canonical representation.

{\textit Example.---}Consider again a system described by the Hamiltonian \eqref{Hamiltonian-1} and parametrization \eqref{alpha-rtp}. The unitary evolution generated by  this Hamiltonian  provides  the canonical mapping of the algebra
onto the group \cite{GilmoreBook2012}.
On the other hand, an arbitrary unitary matrix   $U\in SU(2)$ can be written in a unique way as a product of two group elements  \cite{GilmoreBook2012}
\begin{eqnarray}\label{Ucoset}
U(\eta,\gamma,\xi)=\Omega^{\left( 2\right) }(\gamma,\xi)\Omega^{\left(1\right) }(\eta),
\end{eqnarray}
where $\Omega^{\left(1\right) }(\eta)=\exp\{-i\eta\sigma_z/2\}$  is  diagonal (in the computational basis $\{\ket{0},\ket{1}\}$),   corresponding to the one-dimensional Cartan subalgebra of $su(2)$, and
$\Omega^{\left( 2\right) }(\gamma,\xi)=\exp\{-i\gamma(\sigma_x\sin{\xi}+\sigma_y\cos{\xi})/2\}$ is an arbitrary element of the two-dimensional quotient  space $S^2=SU(2)/U(1)$. The relation
between the canonical  parameters $\{r,\vartheta,\varphi\}$  and the aim parameters $\{\eta,\gamma,\xi\}$ is
\begin{eqnarray}
r&=&\frac{1}{t}\cos^{-1}\left(\cos{\frac{\gamma}{2}}\cos\frac{\eta}{2}\right), \\
\vartheta&=&\cos^{-1}\left(\frac{\cos{\frac{\gamma}{2}}\sin{\frac{\eta}{2}}}{\sqrt{1-\cos^2{\frac{\gamma}{2}}\cos^2{\frac{\eta}{2}}}}\right), \\
\varphi&=&\tan^{-1}\left(\cot\left(\xi+\frac{\eta}{2}\right)\right),
\end{eqnarray}
which  can be used to calculate the transfer matrix $S$. After calculating $S$, and  regarding that we have an explicit expression for $\boldsymbol{m}$-vectors in  the parameters $\{r,\vartheta,\varphi\}$, Eq \eqref{m-R}, one can invoke  Eq. \eqref{m-beta-m-alpha} and get
\begin{eqnarray}\label{m-vectors-eta}
{\boldsymbol{m}_\eta} &=&\frac{1}{2}\left(0, 0, 1 \right)^\T, \\  \label{m-vectors-gamma}
{\boldsymbol{m}_\gamma} &=&\frac{1}{2}\left(\sin\Gamma, \cos\Gamma, 0 \right)^\T, \\ \label{m-vectors-xi}
{\boldsymbol{m}_\xi} &=&\sin{\frac{\gamma}{2}}\left(\cos\frac{\gamma}{2}\cos\Gamma, -\cos\frac{\gamma}{2}\sin\Gamma, -\sin\frac{\gamma}{2} \right)^\T,
\end{eqnarray}
where  $\Gamma=\xi+\eta$. Having  these $\boldsymbol{m}$-vectors in hand,  one can easily use Eq. \eqref{FisherQubit-n} to calculate QFI of the coset parameters for an arbitrary initial state. For instance, when the initial state is diagonal in the computational basis $\{\ket{0},\ket{1}\}$,  we get
\begin{eqnarray}\label{F(c)coset2}
F(\eta,\gamma,\xi)=(p_{0}-p_{1})^{2}\begin{pmatrix}
0&0 & 0\\
0 &1&0 \\
0 &0 &\sin^{2}{\gamma}
\end{pmatrix}.
\end{eqnarray}
This simple form for $F(\eta,\gamma,\xi)$, in particular vanishing $F_{\eta\eta}(\eta,\gamma,\xi)$,  is not surprising as we have  assumed that $\rho_0$ is diagonal  in the Cartan basis of the algebra, so that $\rho$ cannot encode any parameters of the Cartan subalgebra.

\section{Conclusion}
In this paper, we have considered the  quantum Fisher information for  unitary processes with special attention to $SU(2)$ processes. In particular, we have presented a new formulation to calculate QFI matrix in terms of vectors $\boldsymbol{m}_{\theta_i}\in \mathbb{R}^3$, associated  to each estimation parameter $\theta_i$. Our method gives a closed relation for the QFI matrix and reveals, simply,  its features.  Furthermore, for a general Hamiltonian with arbitrary parametrization, we have provided a closed relation to calculate vectors $\boldsymbol{m}_{\theta_i}$. The relation is expressed in terms of derivatives of the Hamiltonian parameters with respect to the  parameters under estimation. As an application we choose angle-axis parameters, both as Hamiltonian parametrization and estimation parameters, and calculate QFI. The generalization of the method to dimensions higher than two is not straightforward and is under further consideration.

Finally, using a linear transformation between two different parameter spaces of a unitary process, we find a relation between QFI matrices of two different classes of estimation parameters. This can be used, in particular, to calculate the QFI of a unitary process in terms of the one  of the same process but with different parametrization, provided that the linear transformation between two parameter spaces is known.
For illustration, we have applied this method for a spin-half system and obtained the QFI matrix of the coset
parameters in terms of the one of the angle-axis parameters.

\acknowledgments The authors  would like to  thank  Fereshte Shahbeigi  for  helpful  discussion and comments. This work was supported by Ferdowsi University of Mashhad under Grant No.  3/44195 (1396/04/17).

\end{document}